\documentclass[twocolumn]{aastex701}
\setcitestyle{numbers}
\bibpunct{[}{]}{,}{n}{}{,}

\usepackage{subcaption}

\begin{document}

\title{PIConGPU modeling of nanoplasma formation in helium nanodroplets irradiated by intense femtosecond laser pulses}

\author[orcid=0000-0000-0000-0001,sname=Medina,gname=Cristian]{Cristian Medina}
\affiliation{Max-Planck-Institut f\"ur Kernphysik, Heidelberg, Germany}
\email[show]{cristian.medina@mpi-hd.mpg.de}

\author[orcid=0000-0001-8965-1149,sname=Steiniger,gname=Klaus]{Klaus Steiniger}
\affiliation{Helmholtz-Zentrum Dresden - Rossendorf, Center for Advanced Systems Understanding (CASUS), 02826 G\"{o}rlitz, Germany}
\email{k.steiniger@hzdr.de}

\author[orcid=0009-0005-6873-8292,sname='Marr\'{e}',gname='Brian Edward']{Brian Edward Marr\'{e}}
\affiliation{Helmholtz-Zentrum Dresden - Rossendorf, Bautzner Landstr. 400, 01328 Dresden, Germany}
\email{b.marre@hzdr.de}

\author[sname=Mudrich,gname=Marcel]{Marcel Mudrich}
\affiliation{Institute of Physics, University of Kassel, Kassel, 34132, Germany}
\email{mudrich@phys.au.dk}

\author[sname={L{\ae}gdsmand},gname={Asbj{\o}rn {\O}.}]{Asbj{\o}rn {\O}. L{\ae}gdsmand}
\affiliation{Institute of Physics, University of Kassel, Kassel, 34132, Germany}
\email{asbjoern.oernemark.laegdsmand@uni-kassel.de}

\author[sname=Krishnan,gname=SivaRama]{SivaRama Krishnan}
\affiliation{Department of Physics and Quantum Center of Excellence for Diamond and Emerging Materials, Indian Institute of Technology Madras, Chennai, 600036 India}
\email{srk.iitmadras@gmail.com}

\author[sname=Sishodia,gname=Keshav]{Keshav Sishodia}
\affiliation{ELI Beamlines facility, The Extreme Light Infrastructure ERIC, Za Radnicí 835, Dolní Břežany, 253 41, Czech Republic}
\email{keshav.sishodia@eli-beams.eu}

\author[sname=Albrecht,gname=Martin]{Martin Albrecht}
\affiliation{ELI Beamlines facility, The Extreme Light Infrastructure ERIC, Za Radnicí 835, Dolní Břežany, 253 41, Czech Republic}
\email{Martin.Albrecht@eli-beams.eu}

\author[sname=Klimešová,gname=Eva]{Eva Klimešová}
\affiliation{ELI Beamlines facility, The Extreme Light Infrastructure ERIC, Za Radnicí 835, Dolní Břežany, 253 41, Czech Republic}
\email{Eva.Klimesova@eli-beams.eu}

\author[sname=Krikunova ,gname=Maria]{Maria Krikunova }
\affiliation{ELI Beamlines facility, The Extreme Light Infrastructure ERIC, Za Radnicí 835, Dolní Břežany, 253 41, Czech Republic}
\affiliation{Wildau Technical University of Applied Sciences, Hochschulring 1, 15745
Wildau, Germany}
\email{Maria.Krikunova@eli-beams.eu}

\author[sname=Andreasson ,gname=Jakob]{Jakob Andreasson }
\affiliation{ELI Beamlines facility, The Extreme Light Infrastructure ERIC, Za Radnicí 835, Dolní Břežany, 253 41, Czech Republic}
\email{Jakob.Andreasson@eli-beams.eu}

\author[sname=Zhang,gname=Weiyu]{Weiyu Zhang}
\affiliation{Max-Planck-Institut f\"ur Kernphysik, Heidelberg, Germany}
\email{Eva.Klimesova@eli-beams.eu}

\author[sname=Mootheril,gname=Deepthy]{Deepthy Mootheril}
\affiliation{Max-Planck-Institut f\"ur Kernphysik, Heidelberg, Germany}
\email{weiyu.zhang@mpi-hd.mpg.de}

\author[sname=Rapp,gname=Nikolas]{Nikolas Rapp}
\affiliation{Max-Planck-Institut f\"ur Kernphysik, Heidelberg, Germany}
\email{nikolas.rapp@mpi-hd.mpg.de}

\author[orcid=0000-0002-9816-3187,sname=Krasnokutskiy,gname=Sergiy]{Serge A. Krasnokutski }
\affiliation{Friedrich Schiller University Jena, Max Planck Institute for Astronomy, Jena, Germany}
\email{sergiy.krasnokutskiy@uni-jena.de}

\author[sname=Moshammer,gname=Robert]{Robert Moshammer}
\affiliation{Max-Planck-Institut f\"ur Kernphysik, Heidelberg, Germany}
\email{robert.mosshammer@mpi-hd.mpg.de}

\author[sname=Pfeifer,gname=Thomas]{Thomas Pfeifer}
\affiliation{Max-Planck-Institut f\"ur Kernphysik, Heidelberg, Germany}
\email{thomas.pfeifer@mpi-hd.mpg.de}

\begin{abstract}
Helium nanodroplets provide a unique and versatile platform for investigating strong-field-driven nanoplasma dynamics. In this work, we present large-scale, GPU-accelerated particle-in-cell simulations using \textsc{PIConGPU} to study the interaction of pure helium nanodroplets containing up to $10^{6}$ atoms with intense near-infrared femtosecond laser pulses, and compare the results with single-shot velocity-map electron imaging and ion measurements. The simulations describe the plasma evolution from the first ionization events to collective electron motion, nanoplasma formation, and early expansion. We show that the calculated electron and ion observables reproduce the main features of the measured spectra in systems with similar cluster sizes and laser intensities. Our results demonstrate that \textsc{PIConGPU} captures the essential physics of nanoplasma formation previously addressed mainly with molecular-dynamics or TDDFT approaches, while remaining computationally efficient and applicable to much larger systems. This establishes \textsc{PIConGPU} as a powerful and scalable tool for connecting nanoplasma theory with experimentally accessible observables.
\end{abstract}


\section{Introduction}
\label{sec:intro}

Nanoplasma formation in clusters driven by intense femtosecond laser pulses provides a well-controlled setting for studying strong-field ionization, plasma formation, and finite-size collective dynamics~\cite{saalmann_mechanisms_2006}. Nanometer-sized clusters and droplets can absorb short laser pulses more efficiently than both isolated atoms and bulk matter. Unlike extended condensed-phase systems, where surface screening can strongly limit light penetration, clusters combine near-solid densities with nearly uniform illumination of the entire target~\cite{pukhov_strong_2002}. Upon irradiation by intense near-infrared (NIR) pulses, clusters can efficiently absorb energy, and evolve into dense, finite plasmas on femtosecond time scales. The resulting dynamics involve a complex interplay of interaction processes—such as tunnel ionization, field-ionization, electron-impact ionization, and collective plasma effects including charge separation, plasma oscillations, and rapid expansion. Because of their finite size and near-solid densities, clusters provide access to collective plasma phenomena in a regime between isolated atoms and extended condensed-phase systems at femtosecond or even picosecond time scales~\cite{fennel_laser-driven_2010,mikaberidze_laser-driven_2009,langbehn_diffraction_2022}.

The homogeneous density, simple electronic structure, and controllable size distribution of helium nanodroplets allow for systematic studies of nanoplasma formation and expansion under reproducible conditions~\cite{mudrich_photoionisaton_2014, bruder_ultrafast_2022}. At the same time, helium’s high ionization potential makes avalanche ionization especially sensitive to seed electrons, laser intensity, and droplet size. This sensitivity has been exploited in previous studies of dopant-induced ignition and nanoplasma dynamics in helium droplets~\cite{mikaberidze_laser-driven_2009,krishnan_dopant-induced_2011,krishnan_evolution_2012,heidenreich_dopant-induced_2017}. More recent experiments have also revealed pronounced shot-to-shot fluctuations in single-shot observables and long-lived modifications of the ignition dynamics following XUV pre-activation~\cite{medina_single-shot_2021,medina_long-lasting_2023,langbehn_diffraction_2022}.

Theoretical descriptions of laser-driven nanoplasmas have traditionally relied on molecular-dynamics (MD) simulations and related microscopic or hybrid nanoplasma models~\cite{peltz_fully_2012,heidenreich_simulations_2007, heidenreich_dopant-induced_2017,molavi_choobini_hydrodynamic-pic_2025, hickstein_observation_2014}. These approaches have successfully reproduced many important experimental signatures and provide detailed insight into strongly coupled plasma dynamics in small clusters. However, fully microscopic MD simulations become increasingly challenging with growing cluster size and are therefore typically restricted to comparatively small systems containing at most a few thousand atoms. In contrast to classical MD, where interactions are resolved microscopically at the particle level, PIC describes the collective electromagnetic response on a grid and couples particles to the self-consistent fields. This coarse-grained treatment makes PIC better suited to large systems and extended simulation volumes, particularly when electromagnetic propagation effects become important.~\cite{peltz_fully_2012, mudrich_photoionisaton_2014,krishnan_evolution_2012}.

Fully electromagnetic particle-in-cell (PIC) methods provide a complementary framework in this regime because they describe the self-consistent evolution of charged particles and electromagnetic fields, including propagation, collective plasma response, and charge separation. They are therefore well suited for studying finite nanoplasmas once electromagnetic propagation effects become relevant~\cite{dawson_particle_1983,verboncoeur_particle_2005}. 

In this work, we use the fully electromagnetic PIC code \textsc{PIConGPU}, including its \textsc{FLYonPIC} atomic physics library~\cite{Burau2010,Bussmann2013,picongpuWWW}, to study nanoplasma formation and early expansion in pure helium nanodroplets containing up to $10^{6}$ atoms. We focus mainly on an initially neutral droplet exposed to an intense NIR pulse, compare its dynamics with a pre-ionized system, and analyze the resulting electron and ion observables in direct comparison with experimentally accessible observables. 
\section{Methods}
\label{sec:methods}

\subsection{Particle-in-cell simulations}

All simulations were performed with the open-source, fully relativistic, electromagnetic, 3D3V particle-in-cell code \textsc{PIConGPU}~\cite{Burau2010, Bussmann2013, picongpuWWW}.  \textsc{PIConGPU} follows the standard explicit electromagnetic PIC method. Charged species (electrons and ions) are represented by macro-particles that evolve in continuous phase space, while electromagnetic fields are defined on a staggered Cartesian grid. Charge and current densities are deposited from the particle ensemble onto the mesh, where Maxwell’s equations are solved self-consistently. The resulting electromagnetic fields are then interpolated back to the particle positions to update their motion.

In our simulations, Maxwell's equations are solved with Yee's method~\cite{yee1966}, particles are advanced using Boris' method~\cite{boris1970}, and currents are deposited using Esirkepov's method~\cite{esirkepov2001}, ensuring discrete consistency with Maxwell’s equations and exact global charge conservation. Electrons ($\mathrm{He}_e$) and ions ($\mathrm{He}_i$) are treated as separate species. Macroparticles use a fourth-order piecewise cubic spline (PQS) shape for charge assignment. Field interpolation from grid to particles employs assigned trilinear interpolation. All simulations are performed in double precision to ensure numerical stability during plasma expansion.

Long-range particle–particle interactions are treated self-consistently through the electromagnetic fields evolved by the PIC cycle. In this approach, charges interact via the grid-resolved electric and magnetic fields rather than direct pairwise Coulomb forces. Short-range collisional processes that are not resolved on the grid scale are accounted for by the Monte Carlo binary collision plugin~\cite{flyonpicwww}, which samples electron–ion and electron–electron interactions locally within a cell. Ionization electrons are created self-consistently as a separate species. Binary Coulomb collisions between all species pairs ($\mathrm{He}_e –\mathrm{He}_e, \mathrm{He}_e–\mathrm{He}_i, \mathrm{He}_i–\mathrm{He}_i$) are included via the relativistic dynamic-log collision operator. 

In addition to Monte Carlo binary collisions, we employ \textsc{PIConGPU}'s atomicPhysics plugin \textsc{FLYonPIC}~\cite{flyonpicwww}.
It tracks the actual atomic state of the ions, including bound-electron excitation.
\textsc{FLYonPIC} is based on averaged states and models the evolution of the atomic state distribution to be self-consistent with the PIC simulation in time and space. The atomic kinetics depend self-consistently on the local plasma conditions without assuming thermal equilibrium.
\textsc{FLYonPIC} is based on the FlyCHK~\cite{chung2007} atomic model and currently implements the following processes: electron-ion collisional de-/excitation, spontaneous deexcitation, electron impact ionization (with Stewart-Pyatt ionization potential depression), autonomous ionization, pressure ionization (according to Stewart-Pyatt ionization potential depression~\cite{stewart1966}), field ionization [BSI + ADK] (with Stewart-Pyatt ionization potential depression).
\textsc{FLYonPIC} solves the atomic rate equation in a time-dependent explicit manner with adaptive sub-stepping of the PIC-cycle and is energy and charge conserving and, in the thermal average, momentum conserving. Thirty bound atomic states of helium are tracked explicitly. Atomic processes are sub-cycled within each PIC time step using an adaptive rate solver with a stability factor $\alpha = 0.1$ and a maximum of 200 atomic sub-steps per PIC step.

The combination of the electromagnetic PIC solver with \textsc{FLYonPIC} and the collisions plugin enables a consistent description of collective plasma dynamics, collisional processes, and atomic kinetics within the nanoplasma. Unlike MicPIC~\cite{peltz_fully_2012}, the present \textsc{PIConGPU} framework does not resolve close charged-particle encounters at atomistic resolution, but its fully electromagnetic, GPU-accelerated implementation with collision and atomic-physics modules makes it well suited to large nanoplasma simulations where field propagation, collective response, and experimentally relevant system sizes must be treated simultaneously.
 
\subsection{Helium nanodroplet model}
The helium nanodroplet is initialized using the $\texttt{SoftSphere}$ density profile implemented in \textsc{PIConGPU}. 
The droplet radius is $R = 25$~nm (50~nm diameter), corresponding to $N \sim 10^{6}$ atoms at liquid helium density. 
The droplet is centered at $(256, 300, 256)$~nm in the simulation domain and embedded in vacuum.

Helium atoms or ions are initialized with a density of $2.1433 \times 10^{28}\,\mathrm{m}^{-3}$ using an FCC profile with random initial positions in each cell and sampling the density with 25 macro-particles per cell per species. The simulation domain is a rectangular box that fully encloses the droplet and a surrounding vacuum region large enough to track emitted electrons and ions, and avoids field reflection from the box edges. We use a Cartesian grid with uniform cell spacing $\Delta x = \Delta y = \Delta z = 0.5$\,nm, chosen such that plasma oscillations are resolved by several tens of time steps $\Delta t=0.958102$\,as and respecting the CFL criterion of the Maxwell solver.

Two scenarios for the initial state are considered:

\begin{enumerate}
\item \textbf{Neutral droplet ignition:} the droplet is initially neutral ($Z=0$), and avalanche ionization develops self-consistently. 
\item \textbf{Pre-seeded nanoplasma:} the droplet is initialized in a preionized configuration where helium atoms are singly ionized, and corresponding electrons are preexisting, mimicking dopant-triggered ignition.

\end{enumerate}

These regimes allow us to disentangle the role of initial charge seeding from strong-field–driven ionization dynamics. 

Both droplets are irradiated by a linearly polarized near-infrared laser pulse with wavelength $\lambda = 800$~nm. The peak intensity is $\sim 10^{15}$~W/cm$^2$ for the initially neutral case and $\sim 10^{14}$~W/cm$^2$ for the singly ionized case. The pulse has a Gaussian temporal envelope with an FWHM duration of 40~fs ($\sigma = 16.986$~fs). The corresponding normalized vector potentials are $a_0 = 0.068$ and $a_0 = 0.0068$, respectively, both well within the nonrelativistic regime ($a_0 \ll 1$). Transversely, a plane-wave profile is used, such that any intensity variation across the droplet arises only from plasma-induced field redistribution.

Field quantities ($\mathbf{E}$, $\mathbf{B}$, $\mathbf{J}$) and derived particle moments (density, energy density, bound-electron density) are recorded every 1000 simulation steps together with the particle output. At each output, particle positions, energies, and emission directions are stored. This enables post-processing of electron energy spectra, ion kinetic-energy distributions, charge-state evolution, and angular emission patterns, allowing direct comparison with the experimentally accessible electron VMI and ion observables. The simulations were carried out using the JUWELS supercomputer at the Jülich Supercomputing Centre, using 16 GPUs in parallel and requiring approximately 36 hours of computational time per run.

\subsection{Experimental methods}
\label{subsec:Exp_methods}

The experiments were performed at the Multipurpose end-station for Atomic, Molecular and Optical Sciences and Coherent Diffraction Imaging (MAC) at the Extreme Light Infrastructure (ELI Beamlines) user facility in Doln\'{\i} B\v{r}e\v{z}any near Prague~\cite{klimesova_multipurpose_2021}. The experimental setup was similar to that used in~\cite{medina_long-lasting_2023}, and the detection scheme followed~\cite{sevaev_studies_2026}. A beam of helium nanodroplets was generated by continuous supersonic expansion of pure He at a stagnation pressure of 50 bar through a 5~$\mu$m  cryogenic nozzle. Under the present expansion conditions, the average droplet size was calculated to be approximately \(2\times10^6\) He atoms per droplet. Droplet sizes were estimated using scaling relations and calibrations reported in~\cite{gomez_sizes_2011}.

An intense near-infrared (NIR) laser pulse at \(800\) nm was focused into the helium droplet beam inside the interaction region of the VMI spectrometer. Charged particles produced by ionization of the helium droplets were detected using a microchannel-plate (MCP) detector coupled to a Timepix camera~\cite{sevaev_studies_2026}. Depending on the spectrometer settings, each laser shot produced a characteristic plasma signal of either electrons or ions. Since a single avalanche-ionized helium droplet can generate a large number of charged particles, the detector was operated at low gain to maintain an approximately linear response despite the large shot-to-shot signal fluctuations.

The electron VMIs were analyzed by fitting the radial intensity distributions with Gaussian profiles, from which characteristic electron energies were extracted. For ions, a direct VMI reconstruction was not possible due to their high kinetic energies and the detector's limited focusing capabilities. Instead, the ion momenta were reconstructed using a free-flight approach. Since the Timepix detector provides both the impact position and the time of arrival of each ion, the initial momentum can be inferred by assuming that the ions originate from the laser focus and propagate in a known extraction field. Because of the limited angular acceptance, only ions traveling toward the detector are recorded. Assuming that the ion energy distribution is approximately symmetric, this measured subset can be taken as representative of the full ion momentum distribution (see Supplementary Material). 
\section{Results}
\label{sec:results}
\paragraph{Nanoplasma formation.}

\begin{figure*}
    \centering
    \includegraphics[width=\textwidth]{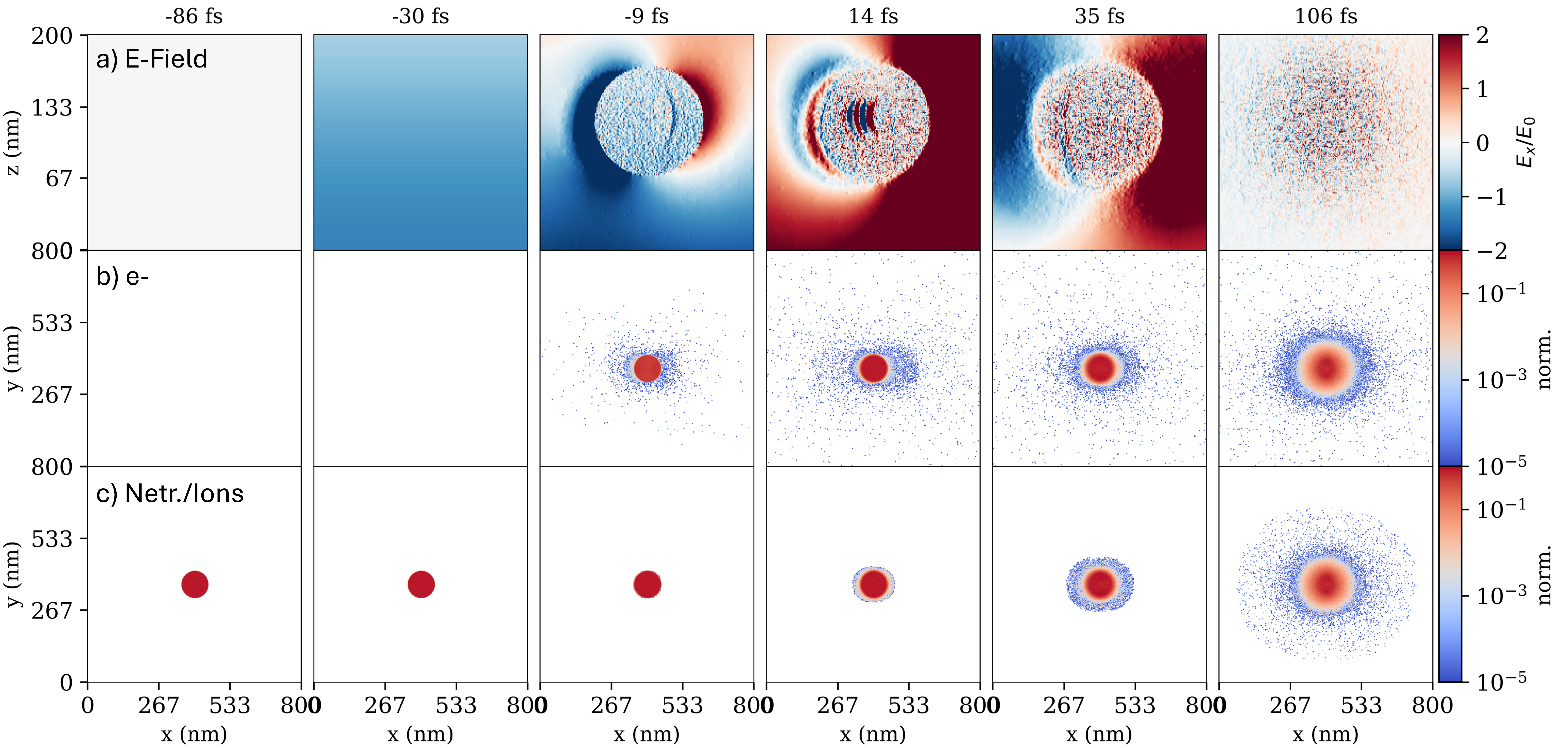}
 \caption{Spatio-temporal evolution of a laser-driven helium nanoplasma for the initially neutral target. Top row: normalized laser field component $E_x/E_0$ in the polarization direction plane (polarization along $x$, propagation along $z$). Middle and bottom row: electron distribution and Helium core-particle distribution (all charge states, $Z=0,1,2$). Times are given relative to the laser peak at the target ($t=0$).}
    \label{fig:maps}
\end{figure*}

Figure~\ref{fig:maps} summarizes the nanoplasma evolution from a neutral helium nanodroplet ($N_{\mathrm{He}}=10^{6}$ atoms) irradiated by an 800 nm laser pulse with a plane-wave transverse profile and peak intensity $I=10^{15}$~W/$\mathrm{cm}^{2}$. It shows time-resolved maps of the normalized laser-field component $E_x/E_0$ (top row), together with projections of the electron charge distribution (middle row) and the helium core-particle density including all charge states (bottom row). The electric-field maps are shown in the $x$--$z$ plane, where the laser polarization is along $x$ and the propagation direction is along $z$, in order to visualize the polarization response of the target during the laser pulse. The electron and ion projections are shown in the $x$-$y$ plane.

Well before the interaction between the HND and the laser pulse ($t=-86$~fs and $t=-30$~fs), the target remains essentially neutral, and no significant response is observed. As the leading edge of the pulse reaches the droplet ($t\approx -20$~fs), the first electrons are generated, and a transient quasi-free electron population rapidly builds up around the droplet. A large fraction of these electrons remains temporarily confined by the space-charge potential, forming a quasi-neutral plasma that shields the cluster interior from the field. The electrons are therefore driven by the laser pulse along the laser polarization axis, producing a pronounced surface-polarized pattern in $E_x/E_0$. This collective motion is characteristic of nanoplasma formation and provides an efficient pathway for energy absorption: quiver-driven electron scattering increases and sustains rapid inner ionization through EII, deepening the cluster potential and further enhancing electron trapping \cite{saalmann_mechanisms_2006,fennel_laser-driven_2010}.

During the main interaction ($t=-9$~fs to $t=35$~fs), in addition to the field modulations near the cluster surface, transient field-enhanced structures appear within the droplet. These wave-like structures correlate with the displacement of the electron cloud: when the electron density is shifted toward one side of the cluster, the strong field modulations are observed on the same side and evolve together with the sub-cycle electron dynamics. 

At later times, after the driving field has vanished, the electron cloud has expanded substantially, and the overall charge density decreases as electrons progressively escape the cluster potential and leave the interaction region. This electron loss leads to a growing net positive charge, consistent with the onset of Coulomb-driven expansion of the helium core. On the present timescale ($\leq 100$~fs), the ionic core remains comparatively localized and expands only weakly compared to the electrons. The observed motion therefore marks only the onset of the expansion, while the main ionic response is expected to develop on longer timescales.

\begin{figure}
    \centering
    \includegraphics[width=\columnwidth]{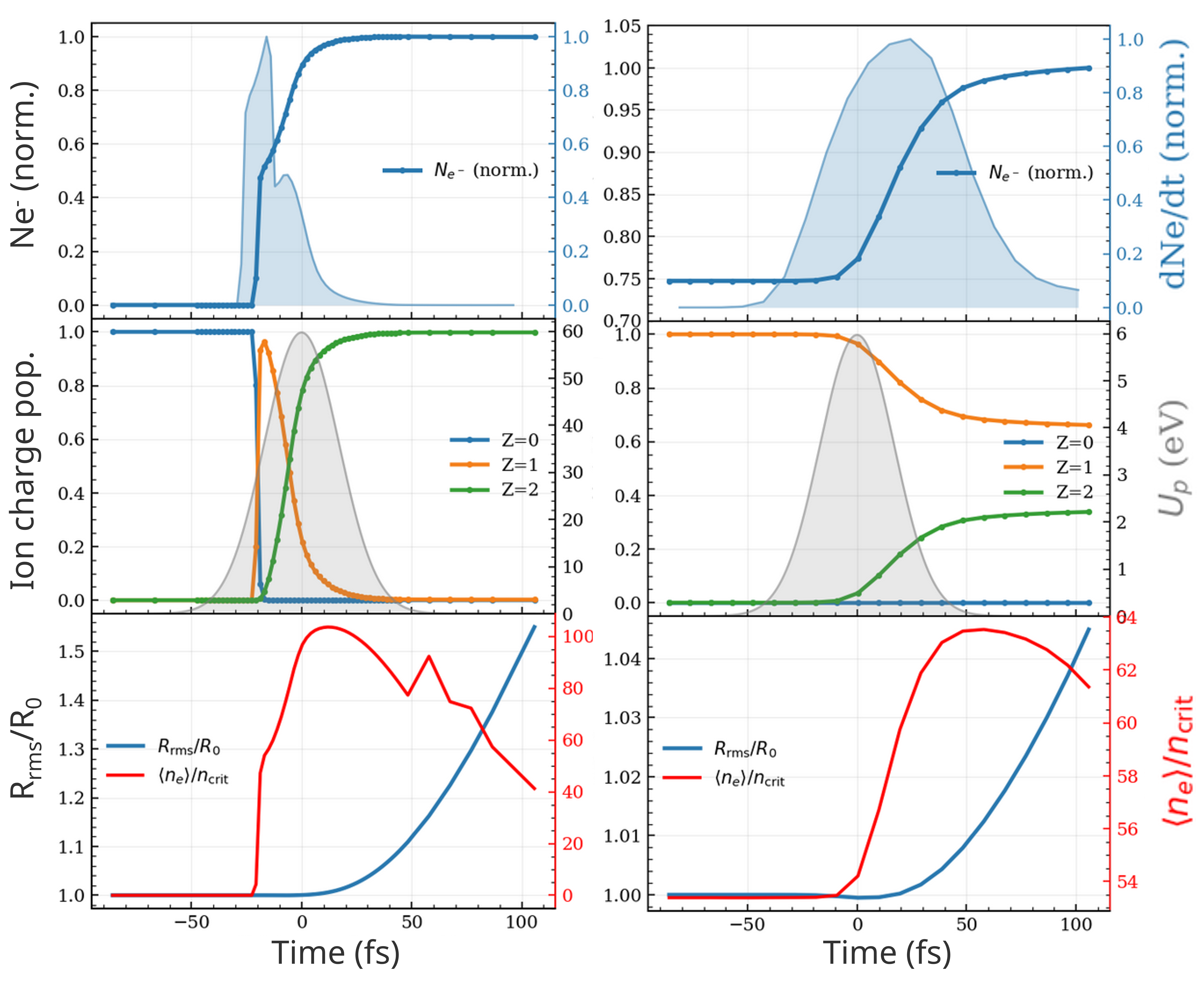}
    \caption{Global dynamics for initially pre-ionized (right) and neutral droplets (left): (top) electron yield $N_e(t)$ and electron generation rate $dN_e/dt$ (filled blue), (middle) ion charge-state fractions with laser intensity envelope (gray), (bottom) expansion metric $R_\mathrm{rms}/R_0$ and mean electron density inside $R_\mathrm{rms}$ normalized to $n_\mathrm{crit}$.}
    \label{fig:dynamics}
\end{figure}

Figure~\ref{fig:dynamics} quantifies the ignition and early expansion dynamics for two initial conditions: the reference case of an initially neutral droplet (left column) and an initially pre-ionized droplet with \(Z=1\) (right column). The neutral and pre-ionized cases are driven at peak intensities of $I_0=1\cdot10^{15}\,$ and $1\cdot10^{14}\mathrm{W\,cm^{-2}}\ $, respectively. The top row shows the quasi-free electron yield $N_e(t)$ (normalized) together with the electron production rate $dN_e/dt$ (filled curve). The middle row reports the charge-state fractions $f_Z(t)$ for $Z=0,1,2$ together with the laser pulse envelope in gray ($t=0$ denotes the pulse peak intensity). The bottom row shows the normalized expansion parameter $R_{\mathrm{rms}}(t)/R_0$ (left axis), where $R_{\mathrm{rms}}$ is the root-mean-square radius of the ion distribution and $R_0$ is the initial radius. The right axis shows the mean electron density inside $R_{\mathrm{rms}}$, $\langle n_e(t)\rangle/n_{\mathrm{crit}}$, evaluated from the quasi-free electron population contained within $R_{\mathrm{rms}}$. Normalization to the critical density $n_{\mathrm{crit}}$ provides a direct indicator of whether the nanoplasma is underdense ($\langle n_e\rangle<n_{\mathrm{crit}}$) or overdense ($\langle n_e\rangle\gg n_{\mathrm{crit}}$) with respect to the driving field.

For the neutral case, the electron yield exhibits a sharp transient increase at the rising edge of the pulse, with a pronounced early peak in $dN_e/dt$ that coincides with the rapid depletion of the neutral population ($Z=0\rightarrow Z=1$) and the onset of ionization throughout the droplet. This stage reflects the generation of early electrons and the subsequent build-up of an ionization avalanche. Close to the pulse maximum, a second, broader contribution to $dN_e/dt$ accompanies the growth of the doubly charged fraction ($Z=2$), indicating continued inner ionization within the already formed nanoplasma. The bottom panel shows that the plasma reaches strongly overdense conditions during the main interaction, with a maximum around $t\approx 40$~fs ($\langle n_e\rangle/n_{\mathrm{crit}}\gg 1$), consistent with the strong screening and surface-localized field response observed in Fig.~\ref{fig:maps}. The expansion, quantified by $R_{\mathrm{rms}}/R_0$, remains weak during the initial ignition stage but becomes more pronounced once substantial net charging and heating have developed, particularly after the build-up of the $Z=2$ population. On the present timescale, this increase marks the onset of Coulomb-driven expansion.

In the pre-ionized case, the early ignition peak associated with the $Z=0\rightarrow Z=1$ transition is absent, while the subsequent dynamics, overdense plasma formation (growth of $Z=2$), and delayed expansion, closely resemble the second, broader contribution dynamics of the neutral case. In this sense, the pre-ionized simulation provides a controlled entry point into the already ignited nanoplasma regime and supports the interpretation of the neutral case in terms of a two-stage evolution. In the following, the analysis therefore focuses mainly on the neutral case.

\begin{figure}

    \centering
    \includegraphics[width=\columnwidth]{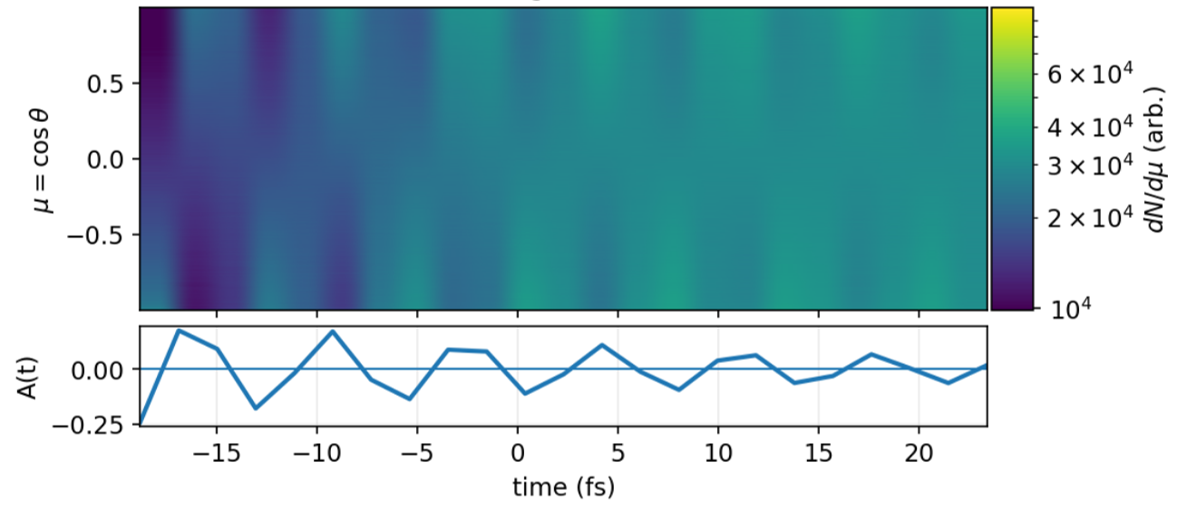}
    \caption{Angular distribution of emitted electrons relative to the laser polarization axis.
    Top: $dN/d\mu$ with $\mu=\cos\theta$. Bottom: anisotropy parameter $A(t)$.}
    \label{fig:angular}
\end{figure}

To visualize the polarization density dynamics of the quasi-free electron cloud observed in Fig.~\ref{fig:maps}, Fig.~\ref{fig:angular} shows the time-dependent angular distribution of the electron population with respect to the laser polarization axis. Here, the distribution in $\mu=\cos\theta$ (with $\theta$ defined for the electron position vector relative to the droplet center) provides a compact measure of the instantaneous polarization of the quasi-free electron cloud along the laser polarization axis. The bottom panel reports the corresponding asymmetry parameter $A(t)=(N_{x>0}-N_{x<0})/(N_{x>0}+N_{x<0})$, the population imbalance between the two hemispheres with respect to the droplet center along $x$.

Upon the initial ionization events ($t\lesssim -10$~fs), the electron distribution becomes strongly anisotropic and $A(t)$ exhibits pronounced oscillations. The modulations are not simply locked to the optical cycle. The electron motion is governed by the combined action of the driving laser field and the restoring force of the transient, positively charged ionic core. These oscillations persist only as long as the system remains approximately quasi-neutral. Around $t\gtrsim 20$~fs, $A(t)$ rapidly relaxes toward zero even while the laser field is still present. This behavior coincides with the onset of electron loss and the breakdown of quasi-neutrality of the plasma: as electrons progressively escape, the net positive charge increases, the electron density decreases, and the balance between electron polarization and ionic restoring fields can no longer be maintained. As a result, this coherent modulation of the electron cloud is quenched and the plasma transitions into the expansion phase, consistent with the subsequent increase of $R_{\mathrm{rms}}/R_0$ in the bottom panel of Fig.~\ref{fig:dynamics}.

\begin{figure}
    \centering
    
    \includegraphics[width=\columnwidth]{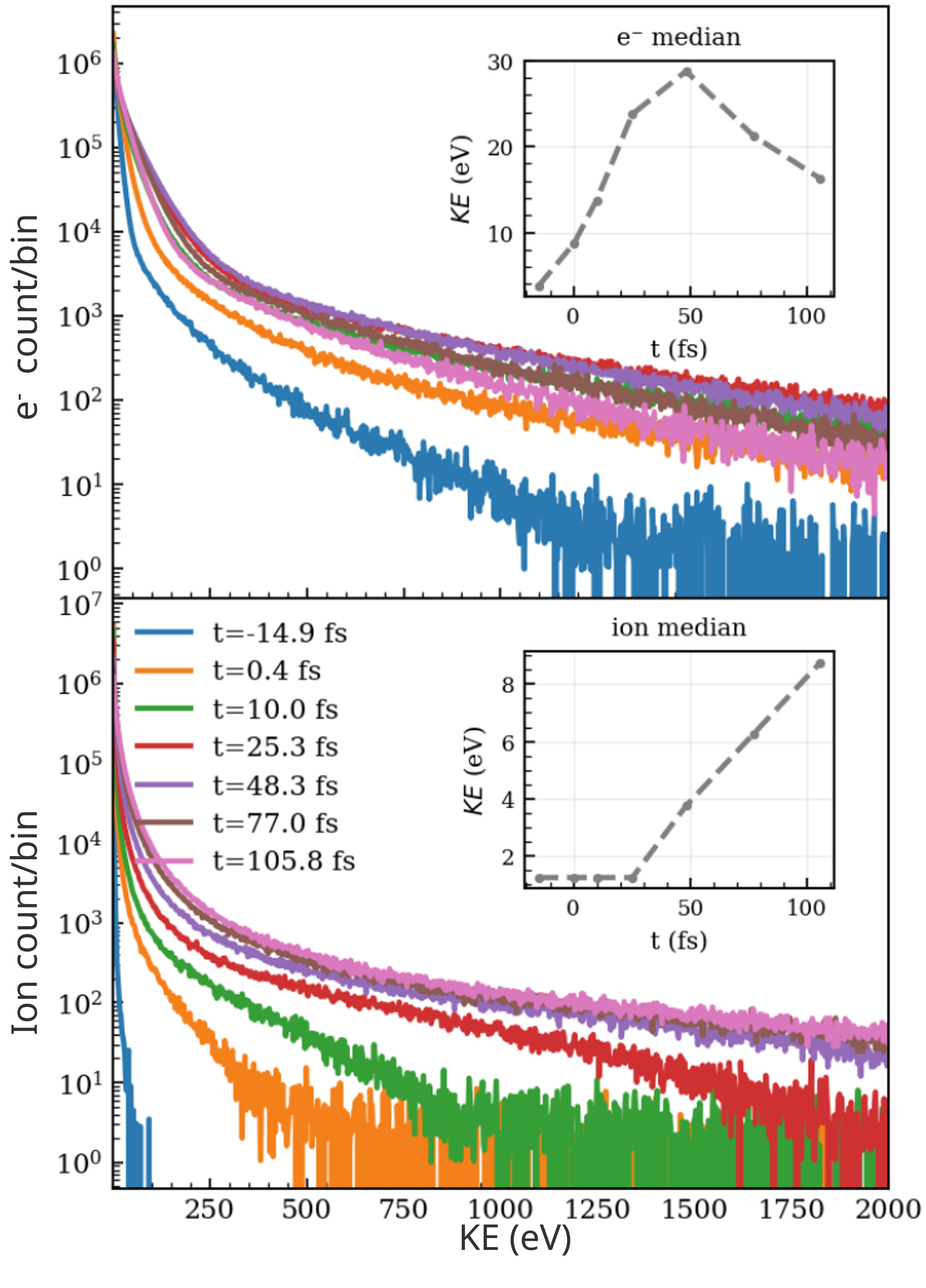}
    \caption{ Time-resolved kinetic-energy distributions of electrons and ions from the PIConGPU simulations. The spectra are shown as counts per energy bin at selected time delays. Insets show the corresponding median kinetic energies as a function of time.}
    \label{fig:energy}
\end{figure}

To connect the simulated nanoplasma dynamics with experimentally accessible observables, we extract the time-resolved kinetic-energy distributions of electrons and ions. Figure~\ref{fig:energy} shows the corresponding energy histograms (counts per bin) at selected time delays. For electrons, the spectrum broadens rapidly during the ignition phase and develops an extended high-energy tail, consistent with efficient heating by the driving field and the forming nanoplasma~\cite{schutte_rare-gas_2014}. The inset shows that the median electron kinetic energy rises quickly, reaches a maximum at \(t \sim 50\)~fs after the laser peak, and then decreases more gradually. This behavior is consistent with rapid energy absorption during the transient overdense stage, followed by cooling as the nanoplasma expands and energetic electrons escape.

In contrast, the ions acquire little kinetic energy during the early, quasi-neutral stage. Only once electrons begin to escape do the ions progressively populate higher kinetic energies. The median ion kinetic energy remains close to zero up to \(t\approx 20\)~fs (see inset) and then increases steadily, consistent with the build-up of net positive charge and the beginning of Coulomb-driven explosion. At later times, a small fraction of ions reaches the keV range, while the median ion energies remain in the few-tens-of-eV range, showing that the high-energy tail accounts for only a minor part of the total yield.

\paragraph{Comparison to experiment.}

\begin{figure}
    \centering
    \includegraphics[width=\columnwidth]{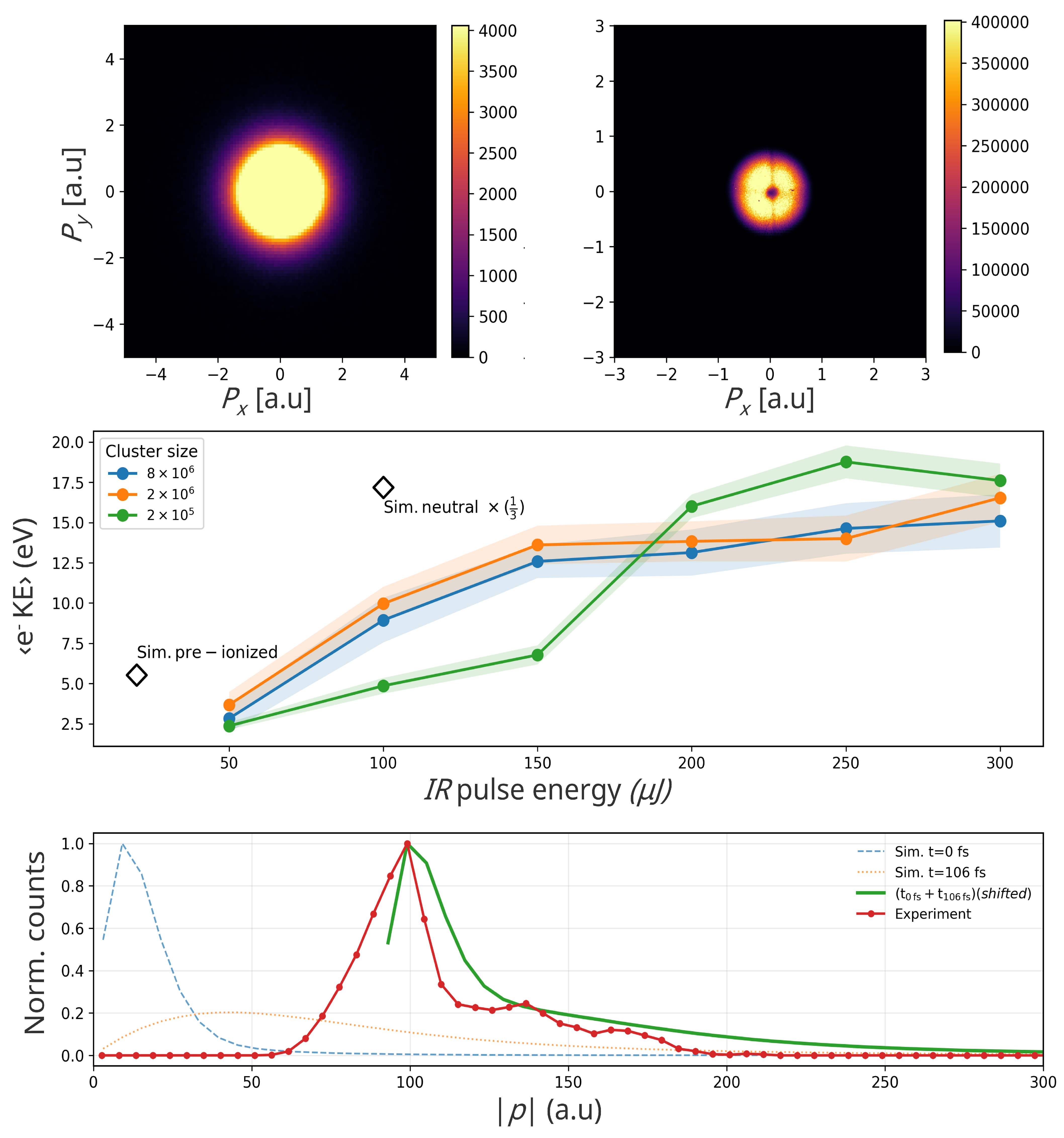}
    \caption{Comparison of PIConGPU simulations with key experimental observables of helium nanoplasmas. Top: simulated and experimental electron VMI images. Middle: average single-shot maximum electron kinetic energy as a function of IR pulse energy for different droplet sizes; the open symbols show analogous values extracted from the simulated VMI images using the same radius-based analysis. Notice that Sim.~Neutral is divided by three for visibility. Bottom: experimental ion momentum distribution (red) compared with the simulated distributions at \(t=0\) (blue) and \(106\) fs (orange), together with their summed distribution, representing the superposition of low- and high-energy nanoplasma contributions (shifted to the experimental peak) for comparison.}
    \label{fig:SimVsExp}
\end{figure}

Figure~\ref{fig:SimVsExp} compares the PIConGPU results with the main experimental observables of the helium nanoplasma. The comparison is intended as a qualitative benchmark of the main features shared by simulation and experiment, rather than a strict one-to-one comparison. It highlights the overall electron VMI morphology, the characteristic electron energies extracted from single-shot images, and the ion momentum distribution.

The top panels compare a simulated electron VMI image at a late extraction time with a representative experimental electron VMI recorded under comparable conditions for pure helium droplets. In both cases, the dominant feature is a compact, nearly isotropic central distribution. The simulation additionally shows a broader contribution extending to larger momenta, consistent with the hotter electron component identified in Fig.~\ref{fig:energy}. In the experiment, part of this more energetic component lies outside the effective detector acceptance and is therefore only partially observed. The central blind spot and the dark vertical stripe in the experimental image are instrumental artifacts caused by a damaged region of the phosphor screen. Overall, the measured morphology is consistent with our earlier single-shot study of dopant-induced helium nanoplasmas~\cite{medina_single-shot_2021}, where molecular-dynamics simulations reproduced similar experimental spectral features. The present agreement indicates that PIConGPU captures the essential plasma-formation dynamics and the main features of the measured electron emission.

The middle panel (b) validates the Timepix-based detection against our previous results and extends the comparison to the simulations in a more systematic way. The experimental points show the average maximum electron kinetic energy  $<KE>$  extracted from single-shot VMI images as a function of pulse energy and droplet size. For each event, this characteristic energy is obtained from the outer radius of the central electron distribution using the radial-based fitting procedure described in the supplementary material. The two open symbols show the corresponding characteristic energies extracted from the simulated VMI images using the same analysis. For visibility, the neutral-run data point is divided by three. The experimental data increase approximately linearly with pulse energy, indicating efficient plasma heating and the emission of more energetic electrons at stronger driving fields as previously seen~\cite {medina_single-shot_2021}. 

The bottom panel shows the ion momentum distribution obtained from the free-flight analysis compared with the simulated distributions at \(t=0\) (blue) and \(106\) fs (orange). In the simulation, the early signal is concentrated at low momenta, while at later times it extends to higher momenta as the Coulomb expansion starts. The red curve represents the experimental ion momentum distribution accumulated from many single-shot events and therefore reflects a broad range of nanoplasma signals, from strongly exploding and energetic events to weaker ones. For a phenomenological comparison, the green curve is obtained by summing the early- and late-time  distributions from the simulation and shifting the result to the experimental peak position. This summed curve is not intended as a prediction for a single event, but rather illustrates how the experimental signal can arise from a superposition of many nanoplasmas with different heating efficiencies and extraction stages. Such a spread is expected because the measured signal combines many single shots with slightly different droplet sizes, ignition conditions, and effective laser intensities. The additional shift can also be understood from the continued ion expansion in the simulation: after the initial plasma formation, the ions keep gaining kinetic energy, so the experimental distribution likely corresponds to a later stage of the dynamics than the earliest simulated extraction time.

Overall, although no one-to-one comparison is possible, the simulations reproduce several of the main experimentally accessible features, including the characteristic VMI distribution and the overall energy and momentum ranges of the emitted particles. The remaining differences are consistent with the experimental averaging over many single-shot nanoplasma events and with detector and extraction-field effects not included in the present model.
\section{Conclusions}

We used the fully electromagnetic PIC code \textsc{PIConGPU} to study nanoplasma formation and early expansion in large helium nanodroplets with experimentally relevant sizes of the order of $10^6$ atoms. The simulations resolve the spatio-temporal evolution from the first ionization events to plasma heating, charge separation, and the onset of Coulomb-driven expansion, and provide a detailed picture of the processes underlying nanoplasma formation and the subsequent electron and ion emission.

By comparing initially neutral and pre-ionized droplets, we found that both evolve toward similar nanoplasma states, but through different ignition pathways. In the neutral case, the droplet first undergoes a rapid ionization stage dominated by field ionization, creating seed electrons, reaching mainly a singly ionized state. This is followed by a slower second stage, where electron-impact ionization builds up an overdense nanoplasma with an increasing doubly ionized helium population, before the system enters a delayed expansion regime. The pre-ionized case provides a controlled entry point into this already singly-ionized regime and shows similar subsequent dynamics at lower laser intensity.

The simulations reproduce several of the main experimental features in both the electron and ion observables. The remaining differences are expected, since the experiment averages over many single-shot nanoplasma events with varying shot-to-shot conditions, whereas the present simulations represent idealized individual cases. Within these limitations, the comparison shows that \textsc{PIConGPU} provides a physically consistent description of the nanoplasma dynamics and a useful connection between the microscopic plasma evolution and the experimentally observed charged-particle emission.

More broadly, the present work highlights the value of \textsc{PIConGPU} for large nanoplasma systems. While more microscopic approaches, such as MD simulations, remain essential for  small clusters and explicitly resolving short-range many-body or quantum effects, they become increasingly difficult to apply in the $>10^6$-atom regime.  In contrast, PIConGPU provides a scalable and computationally efficient framework that retains the collective plasma dynamics, the coupling to realistic laser fields, and effective atomic and collisional physics relevant for this size range. This makes \textsc{PIConGPU} a promising tool for future systematic studies of laser-driven nanoplasmas in large droplets, including more complex phenomena such as shock-wave dynamics within the nanodroplet and doped helium nanoplasma experiments.

\section{Acknowledgments}

The authors acknowledge ELI Beamlines, Dolní Břežany, Czech Republic, for the provided beamtime and thank the facility staff for their assistance and the Institute of Physics of the Czech Academy of Sciences for their support. This work was supported by the project “Advanced research using high-intensity laser-produced photons and particles” (ADONIS) (CZ.02.1.01/0.0/0.0/16-019/0000789), funded by the European Regional Development Fund and the Ministry of Education, Youth and Sports. Part of this research was funded by the Extreme Light Infrastructure ERIC. CM, MM, RM, and TF gratefully acknowledge financial support from the Deutsche Forschungsgemeinschaft (DFG) and from COST Action CA21101 COSY, supported by COST (European Cooperation in Science and Technology). MM gratefully acknowledges support from the Carlsberg Foundation. CM and KS acknowledge the Jülich Supercomputing Centre (JSC) for providing computing time on the JUWELS supercomputer.

\bibliography{PIConGPU_paper}
\bibliographystyle{ieeetr}

\section{Supplementary Material}
\label{sec:SupMat}

\begin{figure}
    \centering
    \begin{subfigure}{0.5\textwidth}
        \centering
        \includegraphics[width=\linewidth]{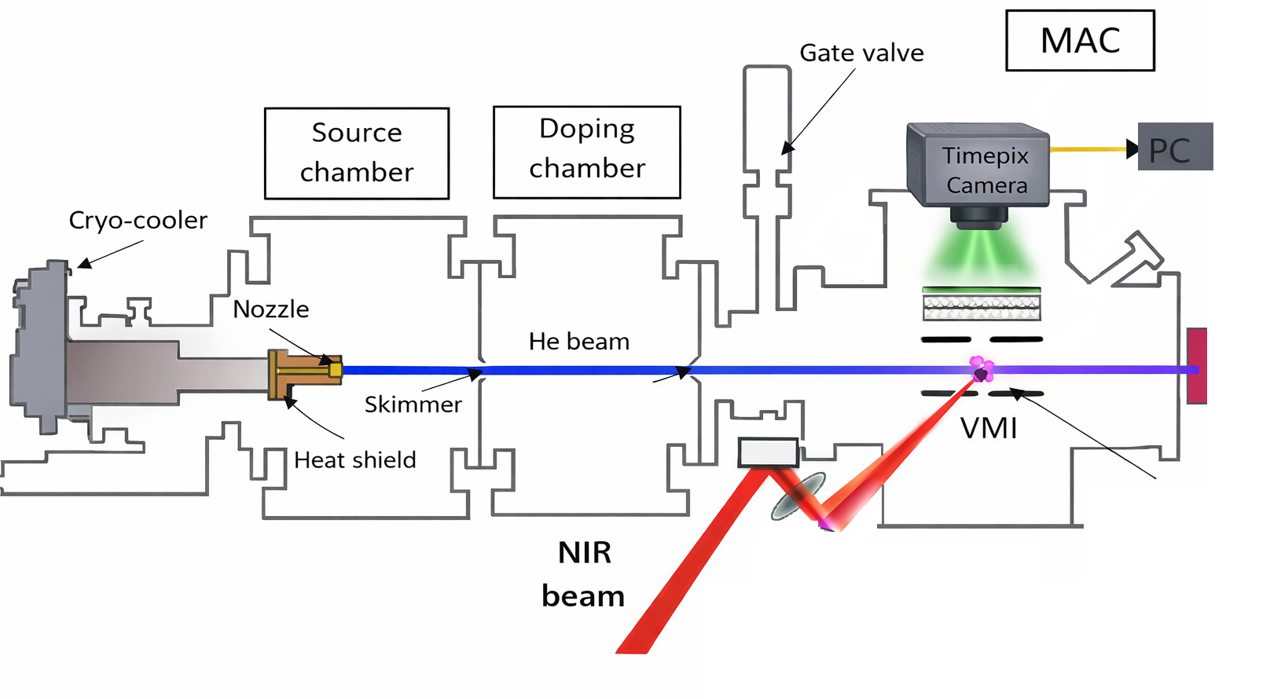}
        \caption*{(a) Experimental setup.}
    \end{subfigure}
    \hfill
    \begin{subfigure}{0.4\textwidth}
        \centering
        \includegraphics[width=\linewidth]{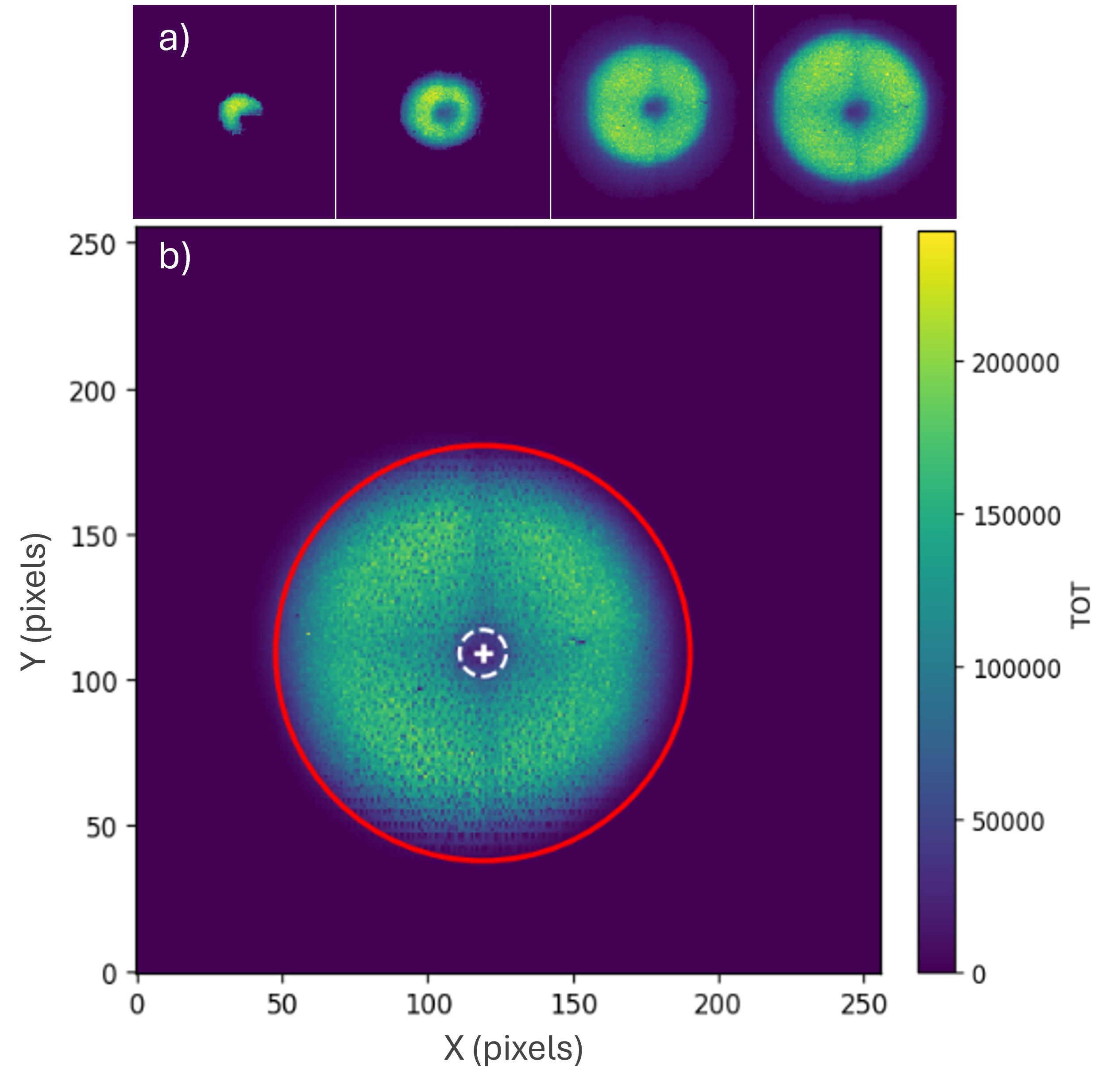}
        \caption*{(b) Representative single-shot VMI analysis.}
    \end{subfigure}
    \caption{Experimental setup and single-shot VMI analysis. Left: schematic of the MAC endstation. A beam of helium nanodroplets is generated in the source chamber and propagated through the skimmers into the interaction region of the VMI spectrometer, where it is intersected by a focused NIR laser pulse. Right: representative single-shot electron VMIs recorded in VMI mode. Panel (a) shows typical single-shot nanoplasma images of increasing size and brightness. Panel (b) shows a representative event and the radial analysis used in this work. The red circle marks the characteristic radius from a Gaussian fit, from which a characteristic maximum electron kinetic energy is extracted.}
    \label{fig:setup_vmi_sup}
\end{figure}

Figure~\ref{fig:setup_vmi_sup} summarizes the experimental arrangement and the single-shot electron-image analysis used in this work. The setup shown on the left is similar to that of Ref.~\cite{medina_long-lasting_2023}, except that here only a single NIR pulse is focused into the interaction region, and the phosphor-screen signal is recorded with a Timepix camera instead of a conventional CCD camera. The right-hand panels show representative single-shot electron VMIs recorded with the Timepix detector. For each nanoplasma event, the radial intensity distribution was analyzed by a Gaussian fit, from which a characteristic maximum electron kinetic energy was extracted (Red circle). These fitted quantities are the basis for the statistical electron-energy analysis presented in the main text.

The VMI spectrometer was calibrated by recording photoelectrons emitted from rare-gas atoms introduced directly into the spectrometer chamber and ionized under well-defined conditions.

Because of the much larger masses and kinetic energies of the ions, a direct VMI reconstruction was not possible. Instead, the ions were analyzed using a free-flight approach: they are accelerated toward the detector by a known homogeneous electric field, and their momenta are reconstructed from the measured impact positions and times of flight.
 For a uniform extraction field \(E\) along the spectrometer axis \(z\), the ion motion satisfies
 
\begin{equation}
L = v_{z,0}\, t_{\mathrm{TOF}} + \frac{1}{2}\frac{qE}{m} t_{\mathrm{TOF}}^2,
\end{equation}
where \(L\) is the flight distance, \(t_{\mathrm{TOF}}\) is the measured time of flight, \(m\) is the ion mass, and \(q\) is its charge state. The initial longitudinal momentum is then
\begin{equation}
p_z = m v_{z,0} = \frac{mL}{t_{\mathrm{TOF}}} - \frac{qE}{2}\, t_{\mathrm{TOF}}.
\end{equation}
The transverse momentum components follow from the measured detector coordinates,
\begin{equation}
p_x = m\frac{x-x_0}{t_{\mathrm{TOF}}}, \qquad
p_y = m\frac{y-y_0}{t_{\mathrm{TOF}}},
\end{equation}
where \((x_0,y_0)\) denotes the projection of the interaction point onto the detector plane. The total ion momentum and kinetic energy are then given by
\begin{equation}
p = \sqrt{p_x^2+p_y^2+p_z^2},
\qquad
E_{\mathrm{kin}} = \frac{p^2}{2m}.
\end{equation}

\end{document}